\documentclass[a4paper, 12pt]{article}
\usepackage{graphicx}
\title{Multi-well potentials and extended supersymmetric quantum mechanics.
}

\author{V.P. Berezovoj \\
A.I. Akhiezer Institute of Theoretical Physics, NSC ''KIPT'', Kharkov\\
e-mail: \texttt{berezovoj@kipt.kharkov.ua}}
\begin{document}

\maketitle
\begin{abstract}
Using the formalism of extended $N=4$ supersymmetric quantum mechanics we consider the procedure of the construction of multi--well potentials. We demostrate the form--invariance of Hamiltonians entering the supermultiplet, using the presented relation for integrals, which contain fundamental solutions. The possibility of partial $N=4$ supersymmetry breaking is determined. We also obtain exact forms of multi--well potentials, both symmetric and asymmetric, using the Hamiltonian of harmonic oscillator as initial. The modification of the shape of potentials due to variation of parameters is also discussed, as well as  application of the obtained results to the study of tunneling processes.
\end{abstract}

\section{Introduction}

Hamiltonians with multi-well potentials are of interest for classical, as well as quantum systems. Among the most interesting applications of multi-well potentials in classical dynamics is the problem of transitions of particles from one local minimum to another under the influence of different types of noise. This problem is common for many physical systems from Universe to microcosm. History of calculations of transition rates from local minimum to another take rise in the famous paper of Kramers \cite{susy-article-ref:1} and after 70 years is still far from completion. Especially interesting is observed effect of amplification of transitions over the barrier under the action of weak time dependent periodic signal, which is called stochastic resonance \cite{susy-article-ref:2, susy-article-ref:3}.

Dynamics in multi-well potentials in quantum mechanical case is not less interesting. That's enough to mention the tunneling effect, which is especially interesting in the connection with wide-ranging research of trapping of atoms of alkali metals in superfluid state. Moreover, when more than one barrier exists, the resonant amplification of tunneling rate is possible, which is known as resonant tunneling \cite{susy-article-ref:4}. Experimental observation of this phenomenon in superconductive heterostructures \cite{susy-article-ref:5} serves as a basis for construction of resonant tunneling diode. It is important to have mechanism to change the parameters of potential(locations of minima, heights of a barriers), to determine the conditions of this phenomenon.

Above mentioned systems, classical as well as quantum, are joined by methods of theoretical analysis. Fokker-Planck equation (FP) is one of the most important kinetic equations, that describe the dynamics of stochastic systems \cite{susy-article-ref:6,susy-article-ref:7}. One of the useful methods for solving FP equation is the eigenfunction expansion method. Due to formal similarity of FP and Shr\"odinger equations, an eigenfunction expansion method, analogous to a bound state expansion treatment of the Schr\"odinger equation, is very useful for the Fokker-Planck equation. In particular, in the case of bistable stochastic system, the knowledge of full set of wave functions and eigenvalues of corresponding quantum Hamiltonian completely determines time evolution of the solutions of FP equation. Moreover, this allows to make a conclusions about the dynamics of corresponding metastable system. Theoretical analysis of the processes in multi-well potentials is complicated due to the fact that existing models operate usually with piecewise potentials (e.g constructed from rectangular or parabolic  wells and barriers), which are far from real potentials. Aside from this, the wave functions and spectrum are unknown in such potentials, which implies the numerical analysis of their properties.

Exact solutions of FP equation in this approach are closely connected with existence of exactly solvable quantum-mechanic problems, which number increases significantly during last years \cite{susy-article-ref:8, susy-article-ref:9}. It is important to mention that the first meaningful example of construction of the significantly nonlinear models of diffusion in bistable system is presented in papers \cite{susy-article-ref:10, susy-article-ref:11}. The formalism of Darboux transformation, used in this papers, have became the one of the most widely used methods of construction of isospectral Hamiltonians in supersymmetric quantum mechanics(SUSY QM). Existence of exactly solvable \cite{susy-article-ref:12} and partially solvable \cite{susy-article-ref:13} quantum mechanical models with multi-well potentials must fill the gap in theoretical analysis of above mentioned processes and serve as a more realistic approximation in their research.

The aim of the present paper, first of all, is to determine the relations between different types of potentials using the framework of the extended supersymmetric quantum mechanics ($N = 4~SUSY~QM$) \cite{susy-article-ref:14, susy-article-ref:15}. Also, we research the possibility of changing parameters of potentials in wide limits and obtaining the exact expressions for spectrum and wave functions. This is especially important when potential has more than one local minimum and, thus, ''resonant'' tunneling is possible. Obtained expressions would be used for derivation of new exactly solvable stochastic models, which is natural extension of previous research in single-well potentials \cite{susy-article-ref:16}.

In the first chapter we give short explanation of $N = 4~SUSY~QM$ structure and procedure of construction of isospectral Hamiltonians with additional states above the ground state of the initial Hamiltonian. We especially emphasis the emergence of multi-well potentials, based on the general properties of the solutions of auxiliary equation of Shr\"odinger type. In particular, we discuss the possibility of partial supersymmetry breaking in $N = 4~SUSY~QM$. In the next chapter we give the expressions, that allows to analyze the obtained potentials and wave functions, regardless of concrete form of initial Hamiltonian. In particular, this allows to generalize the concept of form-invariant potentials \cite{susy-article-ref:17} and calculate in general normalization constants of zero-modes wave functions. Using the harmonic oscillator Hamiltonian as initial, we obtain the expressions of the potentials of isospectral Hamiltonians and wave functions for the case of exact and partially broken supersymmetry. The analysis of conditions for emergence of double and triple-well potentials and ways for varying of their form in wide range is provided in the third chapter. In conclusion we summarize the main results and their possible applications.

\section{Isospectral Hamiltonians of $N = 4~SUSY~QM$ with additional states}

Extended $N = 4~SUSY~QM$ \cite{susy-article-ref:14, susy-article-ref:15} is equivalent of second-order polynomial SUSY QM (reducible case) \cite{susy-article-ref:18, susy-article-ref:19} and assumes the existence of complex operators of sypersymmetries $Q_1$($\bar Q_1$)and $Q_2$($\bar Q_2$), through which the Hamiltonians $H_{\sigma_1}^{\sigma_2}$ could be expressed.

Hamiltonian of $N = 4~SUSY~QM$ has a form($\hbar = m = 1$):
	\begin{equation}
		\begin{array}{c}
		\displaystyle H_{\sigma_1}^{\sigma_2}=\frac{1}{2}(p^2+V_2^2(x)+\sigma_3^{(1)}V'_2(x))\equiv \frac{1}{2}(p^2+V_1^2(x)+\sigma_3^{(2)}V'_1(x)), \\ 
		\displaystyle V_i(x)=W'(x)+\frac{1}{2}\sigma_3^{(i)}\frac{{W''(x)}}{{W'(x)}}
		\end{array}
	\label{susy-article-eqs:11}
	\end{equation}
where $W(x)$ is a superpotential and $\sigma_3^{(i)}$ -- matrices, which commute with each other and have eigenvalues $\pm 1$, $\sigma_3^{(1)}= \sigma_3 \otimes 1, ~\sigma_3^{(2)}= 1 \otimes \sigma_3$. 

Supercharges $Q_i$ of extended supersymmetric quantum mechanics form an algebra: 
	\begin{equation}
		\begin{array}{c}
		\left\{{Q_i, \bar Q_k}\right\}= 2\delta_{ik}H, ~\left\{{Q_i, Q_k}\right\}= \left\{{\bar Q_i, \bar Q_k}\right\}= 0, ~i, k = 1, 2\\
		Q_i = \sigma_-^{(i)}(p+iV_{i+1}(x)), ~\bar Q_i = \sigma_+^{(i)}(p - iV_{i+1}(x))\\ \end{array}
	\label{susy-article-eqs:12}
	\end{equation}
where $V_3(x)\equiv V_1(x), ~\sigma_{_\pm}^{(1)}= \sigma_\pm \otimes 1, ~\sigma_\pm^{(2)}= 1 \otimes \sigma_\pm$.

Hamiltonian and supercharges act on four-dimensional internal space and Hamiltonian is diagonal on vectors $\psi_{\sigma_1}^{\sigma_2}(x, E)$,  where $\sigma_1, ~\sigma_2$ -- eigenvalues of $\sigma_3^{(1)}, ~\sigma_3^{(2)}$. Supercharges $Q_i$($\bar Q_i$) act as lowering (raising) operators for indexes $\sigma_1, ~\sigma_2$. It is convenient to represent the structure of Hamiltonian and connection between wave functions in diagram form:

\begin{figure*}[h]
\begin{center}
 \includegraphics{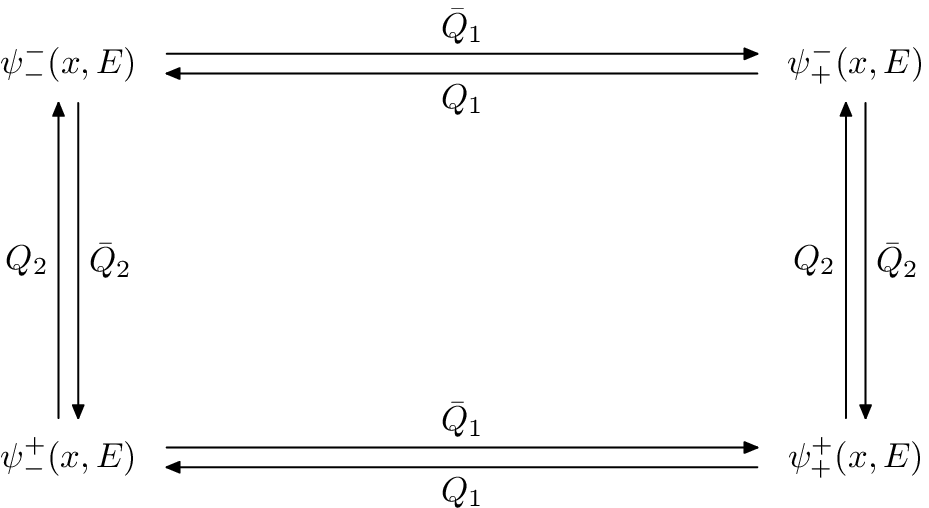}
\end{center}
\end{figure*}

Obviously,  due to commutativity of operators $Q_i$ and $\bar Q_i$ with Hamiltonian, all $\psi_{\sigma_1}^{\sigma_2}(x, E)$ are eigenfunctions of Hamiltonian with the same eigenvalue $E$. The only exception is the case, when wave functions turn to $0$ under the action of generators of supersymmetry.

Construction of isospectral Hamiltonians in the framework of $N = 4~SUSY~QM$ is based on the fact that four Hamiltonians are combined into supermultiplet $H_{\sigma_1}^{\sigma_2}$. Nevertheless, it should be noted that due to symmetry $\sigma_1 \leftrightarrow \sigma_2 - H_{\sigma_1}^{-\sigma_2}\equiv H_{-\sigma_1}^{\sigma_2}$ only three of them are nontrivial. Let's note, that such relation in the case of higher-derivative $SUSY$ \cite{susy-article-ref:19} establishes the correspondence between quasi-Hamiltonian and operators of Shr\"odinger type and is identical for any superpotentials. The procedure of construction of isospectral Hamiltonians, when ground state is removed from the initial Hamiltonian, is treated in \cite{susy-article-ref:20} in detail. We will consider the construction of isospectral Hamiltonians by addition of states above the ground state of initial Hamiltonian. Similar procedure was already performed (e.g. in \cite{susy-article-ref:20}), but the distinctive feature of present research is the obtaining of general results without specification of concrete form of initial Hamiltonian.

Let's consider the auxiliary equation:
	\begin{equation}
		H\varphi(x)= \varepsilon \varphi(x)
	\label{susy-article-eqs:13}
	\end{equation}
As initial let's take one of the Hamiltonians 
	\begin{equation}
		\begin{array}{c}
		H_{\sigma_1}^{\sigma_2}= \frac{1}{2}(p - i\sigma_1 V_2(x))(p+i\sigma_1 V_2(x))+\varepsilon \equiv \\
		\equiv\frac{1}{2}(p - i\sigma_2 V_1(x))(p+i\sigma_2 V_1(x))+\varepsilon \\
		\end{array}
	\label{susy-article-eqs:14}
	\end{equation}
where $\varepsilon$ -- the so-called factorization energy. Hereinafter the energy is measured from $\varepsilon$. Strictly speaking, the supersymmetry relations with supercharges of the form (\ref{susy-article-eqs:12}) and expressed through superpotential $W(x)$ are satisfied not by Hamiltonian $H$ itself, but by a shifted by an amount of $\varepsilon$ Hamiltonian $H - \varepsilon$. However, due to commutativity of supercharges with constant $\varepsilon$, the relations between wave functions of $H$ and $H - \varepsilon$ are the same. When fixing operator $H_{\sigma_1}^{\sigma_2}$, the form of $W(x)$ depends on the choice of factorization energy $\varepsilon$, thereby the Hamiltonians $H_{-\sigma_1}^{\sigma_2}$, $H_{\sigma_1}^{-\sigma_2}$, $H_{-\sigma_1}^{-\sigma_2}$ also have nontrivial dependence on $\varepsilon$.

When $\varepsilon < E_0$ (where $E_0$ is the ground state energy of initial Hamiltonian), the auxiliary equation (\ref{susy-article-eqs:13}) has two linear independent solutions $\varphi_i(x, \varepsilon), ~i = 1, 2$, which are nonnegative and have the following asymptotic behaviour \cite{susy-article-ref:20a}: when $x \to -\infty\;\;\varphi_1(x)\to+\infty \;(\varphi_2(x)\to 0)$, and when $x \to+\infty \;\;\varphi_1(x)\to 0\;(\varphi_2(x)\to+\infty)$, i.e. with appropriately chosen constants the general solution has the form $\varphi(x, \varepsilon, c)= N(\varphi_1(x, \varepsilon)+c\varphi_2(x, \varepsilon))$ ($N$ -- normalization constant) and has no zeros on all axis. Thus, the function $\displaystyle\tilde \varphi(x, \varepsilon, c)= \frac{{N^{- 1}}}{{\varphi(x, \varepsilon, c)}}$ is finite and could be normalized at every concrete choice of $\varepsilon$ and $c$. Let's note that with concrete values of $\varepsilon$ and $c$ $\varphi(x, \varepsilon, c)$ could have local extrema. In this case the natural choice of initial Hamiltonian is $H_-^+$ or $H_+^-$ (which are identical due to symmetry of $H_{\sigma_1}^{\sigma_2}$ under $\sigma_1\leftrightarrow \sigma_2$). Superpotential in this case has the form:
	\begin{equation}
		W(x, \varepsilon, \lambda)= - \frac{1}{2}\ln(1+\lambda \int\limits_{x_i}^x{dt}\tilde \varphi^2(t, \varepsilon, c))
	\label{susy-article-eqs:15}
	\end{equation}
where $\lambda, x_i$ -- two new arbitrary parameters, but one of them is inessential, because it gives an additional contribution to $W(x)$. All Hamiltonians that form a supermultiplet have nontrivial dependence on these parameters.

To consider the connection between Hamiltonians from supermultiplet, let's take $H_+^-$ as initial. Denoting as $\psi_+^-(x, E)$ the solution of equation
	\begin{equation}
		H_+^- \;\psi_+^-(x, E)= E\psi_+^-(x, E)
	\label{susy-article-eqs:16}
	\end{equation}
and using the first representation of Hamiltonian $H_-^-$, we obtain the following relation between, $\psi_-^-(x, E)$ and initial expressions:
	\begin{equation}
		\begin{array}{c}
		\displaystyle H_-^- = H_+^-+\frac{{d^2}}{{dx^2}}\ln \tilde \varphi(x, \varepsilon, c),\\
		\displaystyle  \psi_-^-(x, E_i)= \frac{1}{{\sqrt{2(E_i - \varepsilon)}}}\frac{{W\left\{{\psi_+^-(x, E_i), \varphi(x, \varepsilon, c)}\right\}}}{{\varphi(x, \varepsilon, c)}}\\
		\displaystyle  \psi_-^-(x, E = 0)= \frac{{N^{- 1}}}{{\varphi(x, \varepsilon, c)}}= \tilde \varphi(x, \varepsilon, c)
		\end{array}
	\label{susy-article-eqs:17}
	\end{equation}
The new state with $E = 0$ (with energies measured from $\varepsilon $) appears in Hamiltonian, which by definition has normalized wave function. Normalization of wave functions of excited states is preserved (for discrete spectrum).

Using the second representation $\displaystyle  H_{\sigma_1}^{\sigma_2}= \frac{1}{2}(p - i\sigma_2 V_{\sigma_1}(x))(p+i\sigma_2 V_{\sigma_1}(x))+\varepsilon$ and identity $H_+^- \equiv H_-^+$ the relation between $H_+^+$, $\psi_+^+(x, E)$ and initial Hamiltonian could be obtained:
	\begin{equation}
		\begin{array}{c}
		\displaystyle H_+^+= H_+^- + \frac{{d^2}}{{dx^2}}\ln \left({\frac{{\tilde \varphi(x, \varepsilon, c)}}{{1+\lambda \int\limits_{x_i}^x{dt\;\tilde \varphi^2(t, \varepsilon, c})}}}\right)\;\quad,\\
		\displaystyle \psi_+^+(x, E = 0)= \frac{{N_\lambda^{- 1}\tilde \varphi(x, \varepsilon, c)}}{{\;(1+\lambda \int\limits_{x_i}^x{dt\;\tilde \varphi^2(t, \varepsilon, c))}}}\\
		\displaystyle \psi_+^+(x, E_i)= \frac{1}{{\sqrt{2(E_i - \varepsilon)}}}\left({\frac{d}{{dx}}+\frac{d}{{dx}}\ln \frac{{\tilde \varphi(x, \varepsilon, c)}}{{\;(1+\lambda \int\limits_{x_i}^x{dt\;\tilde \varphi^2(t, \varepsilon, c))}}}}\right)\psi_+^-(x, E_i)
		\end{array}
	\label{susy-article-eqs:18}
	\end{equation}
It is important to mention that normalization of wave function $\psi_+^+(x, E =0)$, as in the case of one-well potentials, could always be performed at any $\tilde \varphi(x, \varepsilon, c)$ and $\lambda$ using the following expression in normalization condition:
	$$
	\displaystyle  \frac{{N_\lambda^{- 2}\tilde \varphi^2(x, \varepsilon, c)}}{{(1+\lambda N^{- 2}\int\limits_{- \infty}^x{dt\tilde \varphi^2(t, \varepsilon, c)})^2}}= - \frac{{N_\lambda^{- 2}}}{{\lambda N^{- 2}}}\frac{d}{{dx}}\frac{1}{{(1+\lambda N^{- 2}\int\limits_{- \infty}^x{dt\tilde \varphi^2(t, \varepsilon, c)})}}
	$$
which easily gives the relation between normalization constants: $N_\lambda^{-2}=(1+\lambda)N^{-2}$. Normalization of $\psi_+^+(x, E_i)$ is the same as for $\psi_+^-(x, E_i)$ at any choice of $\tilde \varphi(x, \varepsilon, c)$. The usage of superpotential (\ref{susy-article-eqs:15}) with $\tilde \varphi(x, \varepsilon, c)$ corresponds to exact supersymmetry, which leads to existence of zero-modes in $H_-^-$ and $H_+^+$. Existence of two zero-modes in super Hamiltonian of $N = 4~SUSY~QM$ is caused by the fact that Witten theorem \cite{susy-article-ref:21} modifies when entanglement conditions are nonlinear (unlike the case of $N = 2~SUSY~QM$) as discussed in detail in \cite{susy-article-ref:18}.

Let's consider the case when expression (\ref{susy-article-eqs:15}) contains one of the particular solutions, e.g. $\varphi_1(x, \varepsilon)$, instead of $\varphi(x, \varepsilon, c)$. If one of the particular solutions of second order equation is known, the second solution could be calculated from the relation $\displaystyle \varphi_2(x, \varepsilon)= \varphi_1(x, \varepsilon)\int\limits_{-\infty}^x{dt\frac{1}{{\varphi_1^2(t, \varepsilon)}}}$. Thus, superpotential (\ref{susy-article-eqs:15}) obtains the form:
	\begin{equation}
	\begin{array}{c}
		\displaystyle W(x, \varepsilon, \lambda)= - \frac{1}{2}\ln(1+\lambda \int\limits_{- \infty}^x{dt}\;\frac{1}{{\varphi_1^2(t, \varepsilon)}})\equiv \\
		\displaystyle \equiv \frac{1}{2}\ln \left({\frac{{\varphi_1(x, \varepsilon)}}{{\varphi_1(x, \varepsilon)+\lambda \varphi_2(x, \varepsilon)}}}\right)= \frac{1}{2}\ln \left({\frac{{\varphi_1(x, \varepsilon)}}{{\varphi(x, \varepsilon, \lambda)}}}\right)
	\end{array}
	\label{susy-article-eqs:19}
	\end{equation}
It is easy to show, that the state with energy $E = 0$ in the spectrum of $H_-^-$ is absent (i.e., the wave function $\psi_-^-(x, E = 0)$ is nonnormalizable) and thus the spontaneously broken super symmetry exists. On the one hand, the state with zero energy and wave function $ \displaystyle \psi_+^+(x, E = 0)\sim \frac{1}{{\varphi(x, \varepsilon, \lambda)}}$, which could be normalized at certain values of $\lambda > 0$, appears in the spectrum of $H_+^+$. In this case the exact $N = 2$ supersymmetry takes place. From the other hand, it is known \cite{susy-article-ref:21, susy-article-ref:22}, that partial supersymmetry breaking is impossible in $N = 4~SUSY~QM$ without introduction of central charges. This contradiction resolves, when taking into consideration that employment of factorization energy $\varepsilon $ in construction of isospectral Hamiltonians is the simplest way of incorporation of central charges in $N = 4~SUSY~QM$. The similar situation occurs in consideration of form--invariant potentials \cite{susy-article-ref:23}. More complete and consistent consideration of $N = 4~SUSY~QM$ with central charges is given in \cite{susy-article-ref:24}, but this is outside the scope of the current paper.
 
\section{General properties of multi-well potentials}

We will start the consideration of properties of isospectral Hamiltanians, derived in previous chapter without concretization of the from of initial Hamiltonian, from obtaining the helpful mathematical expressions. Let $y_1$ and $y_2$ be two linear independent solutions of homogeneous second order equation, then the following condition is hold: 
	\begin{equation}
	\begin{array}{c}
		\displaystyle\int\limits_{x_i}^x {W\{y_1, y_2\} \over\left(A_1 y_1(t)+A_2 y_2(t)\right)^2}dt = - {1\over A_1^2+A_2^2} \left[\left({A_2 y_1(x)- A_1 y_2(x)\over A_1 y_1(x)+A_2 y_2(x)}\right)-\right. \\
		\displaystyle \left.- \left({A_2 y_1(x_i)- A_1 y_2(x_i)\over A_1 y_1(x_i)+A_2 y_2(x_i)}\right)\right]
  	\end{array}
	\label{susy-article-eqs:21}
	\end{equation}
where $W\{y_1, y_2\}= y_1 y'_2 - y'_1y_2$ is Wronskian, which is independent on $x$ for the second order equation, reduced to canonical form, as Shr\"odinger equation, and thus could be passed from integral. The special from of this relation for the parabolic cylinder functions was obtained in \cite{susy-article-ref:10}. This relation is very useful for calculation of the integrals in the above given expressions.

Let's consider the case of $c = 1$ in the general solution of auxiliary equation (\ref{susy-article-eqs:14}). This allows to avoid the unnecessary awkwardness, but nevertheless, reveal the fundamental features of Hamiltonians and wave functions of $N = 4~SUSY~QM$, using only general properties of the solutions of auxiliary equation when $\varepsilon < E_0$. First of all, it is natural to put $x_i = - \infty$ in expression (\ref{susy-article-eqs:15}) of superpotential, because $ \displaystyle \tilde \varphi(x, \varepsilon, c = 1)= \frac{{N^{- 1}}}{{\varphi_1(x, \varepsilon)+\varphi_2(x, \varepsilon)}}$ tends to $0$ in this limit and function with asymptotic $\displaystyle{1\over\varphi_i(x, \varepsilon)}\to 0$ when $x \to - \infty $ always exists, if using particular solution $\varphi_i(x, \varepsilon)$.

	\begin{equation}
		\begin{array}{c}
		\displaystyle N^{- 2}\int\limits_{- \infty}^x{\frac{{dt}}{{\left({\varphi_1(t, \varepsilon)+c\;\varphi_2(t, \varepsilon)}\right)^2}}}=\\
		\displaystyle = - \frac{{N^{- 2}}}{{\;(1+c^2)\;W\{\varphi_1, \varphi_2 \}}}\left[{\Delta(x, \varepsilon, c)- \Delta(- \infty, \varepsilon, c)}\right] \\ 
		\displaystyle N^{- 2}= - \frac{1}{{\;(1+c^2)\;W\{\varphi_1, \varphi_2 \}}}\left[{\Delta(+\infty, \varepsilon, c)- \Delta(- \infty, \varepsilon, c)}\right] \\
		\displaystyle \Delta(x, \varepsilon, c)= \frac{{c\varphi_1(x, \varepsilon)- \varphi_2(x, \varepsilon)}}{{\varphi_1(x, \varepsilon)+c\varphi_2(x, \varepsilon)}}
		\end{array}
	\label{susy-article-eqs:22}
	\end{equation}
Using above considered relations, it is easy to obtain the expression for superpotential up to constant term when $c = 1$: 
	\begin{equation}
		\begin{array}{c}
		\displaystyle W(x, \varepsilon, \lambda)= - \frac{1}{2}\ln \left({1+\lambda N^{- 2}\int\limits_{- \infty}^x{\frac{{dt}}{{\left({\varphi_1(x, \varepsilon)+\varphi_2(x, \varepsilon)}\right)^2}}}}\right)=\\
		\\
		\displaystyle = - \frac{1}{2}\ln \left({\frac{{\varphi_1(x, \varepsilon)+\Lambda(\varepsilon, \lambda)\varphi_2(x, \varepsilon)}}{{\varphi_1(x, \varepsilon)+\varphi_2(x, \varepsilon)}}}\right)\\
		\\
		\displaystyle \Lambda(\varepsilon, \lambda)= \frac{{\Delta(\infty, \varepsilon, c = 1)- \lambda -(\lambda+1)\;\Delta(- \infty, \varepsilon, c = 1)}}{{\Delta(\infty, \varepsilon, c = 1)+\lambda -(\lambda+1)\;\Delta(- \infty, \varepsilon, c = 1)}}
		\label{susy-article-eqs:23}
		\end{array}
	\end{equation}
Let's note that $\Delta(\pm\infty, \varepsilon, \lambda)$, which enter the $\Lambda(\varepsilon, \lambda)$, are determined by the asymptotic behavior of solutions of auxiliary equation. Due to this, while in the case of $H_-^-$ potential is determined by symmetric combination $\varphi_1(x, \varepsilon)+\varphi_2(x, \varepsilon)$, in the case of $H_+^+$ -- by asymmetric $\varphi_1(x, \varepsilon)+\Lambda(\varepsilon, \lambda)\;\varphi_2(x, \varepsilon)$:
	\begin{equation}
		H_-^-= H_+^- - \frac{{d^2}}{{dx^2}}\ln(\varphi_1(x, \varepsilon)+\varphi_2(x, \varepsilon))
	\label{susy-article-eqs:24}
	\end{equation}
	\begin{equation}
		H_+^+= H_+^- - \frac{{d^2}}{{dx^2}}\ln \left({\varphi_1(x, \varepsilon)+\Lambda(\varepsilon, \lambda)\;\varphi_2(x, \varepsilon)}\right)
	\label{susy-article-eqs:25}
	\end{equation}

In some sense, potentials $U_-^-(x, \varepsilon)$ and $U_+^+(x, \varepsilon, \lambda)$ are form-invariant \cite{susy-article-ref:15}, i.e. potentials and wave functions transform to each other by changing of parameters and their spectra are identical and this holds independently of the choice of initial Hamiltonian. As it would be shown in the next chapter, if potential in $H_-^-$ is multi-well symmetrical potential, then for $H_+^+$ it is asymmetrical. Moreover, the form could be changed when varying $\varepsilon$, as well as $\lambda$. Relations (\ref{susy-article-eqs:22}) and (\ref{susy-article-eqs:23}) are useful also for derivation of exact form of wave functions $\psi_-^-(x, E)$ and $\psi_+^+(x, E)$. Thus, expression for wave functions $\psi_+^+(x, E)$ results from similar expression for $\psi_-^-(x, E)$ by the substitution $\varphi_1(x, \varepsilon)+\varphi_2(x, \varepsilon)\to \left({\varphi_1(x, \varepsilon)+\Lambda(\varepsilon, \lambda)\;\varphi_2(x, \varepsilon)}\right)$. In particular, the normalization constant for $\psi_+^+(x, E = 0)$ could be easily obtained from corresponding expression for $\psi_-^-(x, E = 0)$. To demonstrate the possibility of partial symmetry breaking in $N = 4~SUSY~QM$ we need to show that wave function $\displaystyle \psi_+^+(x, E = 0)\sim \frac{1}{{\varphi(x, \varepsilon, \lambda)}}$ is normalizable. Value of normalization constant could be derived from (\ref{susy-article-eqs:22}) when $c = \lambda > 0$. Later we will return to the calculation of this constant for concrete form of initial Hamiltonian. The potential in $H_+^+$, in turn, has the same form as corresponding potentials in the case of exact supersymmetry with substitution $\Lambda(\varepsilon, \lambda)\to \lambda$.

\section{Isospectral Hamiltonians with almost equidistant spectrum}

To construct concrete expressions for potentials and wave functions we will choose the Hamiltonian with harmonic oscillator ($HO$) potential as initial. Let's consider the solution of auxiliary equation for $\displaystyle \varepsilon<E_0 = \frac{\omega}{2}, (\hbar=m=1)$: 
	\begin{equation}
		\left({\frac{{d^2}}{{dx^2}}+2(\varepsilon - \frac{{\omega^2 x^2}}{2})}\right)\varphi(x, \varepsilon)= 0
	\label{susy-article-eqs:31}
	\end{equation}
Introducing dimensionless variables $\xi = \sqrt{2\omega}x$ we obtain the equation for $\varphi(\xi, \bar\varepsilon)$, where $\displaystyle\bar \varepsilon = \frac{\varepsilon}{\omega}$: 
	\begin{equation}
		\left({\frac{{d^2}}{{d\xi^2}}+(\nu+\frac{1}{2}- \frac{{\xi^2}}{4})}\right)\;\varphi(\xi, \bar \varepsilon)= 0\;, \;\nu = - \frac{1}{2}+\bar \varepsilon
	\label{susy-article-eqs:32}
	\end{equation}

\textbf{a)} \textbf{Exact $N=4~SUSY$}

\begin{figure}
\begin{center}
 \includegraphics[width=0.5\linewidth]{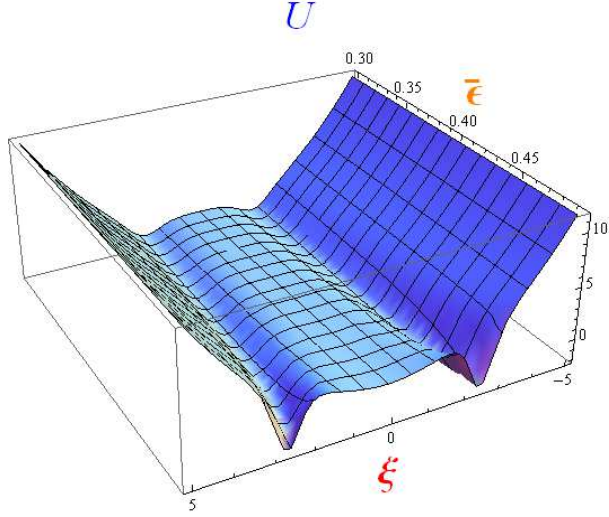}
  \caption{Form of the potential $U_-^-(\xi, \bar \varepsilon, \lambda = 0)$.\label{susy-article-fig:1}}
\end{center}
\end{figure}

\begin{figure}
\begin{center}
 \includegraphics[width=0.5\linewidth]{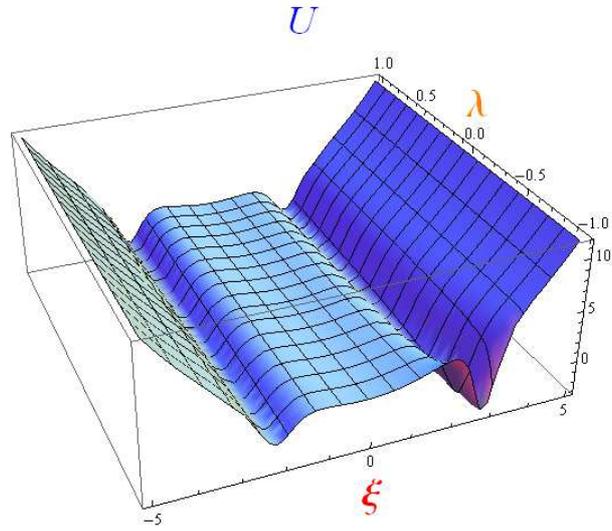}
  \caption{Potential $U_+^+(\xi, \bar \varepsilon = 0.47, \lambda)$.\label{susy-article-fig:2}}
\end{center}
\end{figure}

This equation has two linear independent solutions: $D_\nu(\sqrt 2 \xi), ~D_\nu(-\sqrt 2 \xi)$ -- parabolic cylinder functions. According to terminology, introduced above, we denote $\varphi_1(\xi, \bar\varepsilon)= D_\nu(\sqrt 2 \xi), ~\varphi_2(\xi, \bar\varepsilon)= D_\nu(-\sqrt 2 \xi)$ and Wronskian $\displaystyle W\{\varphi_1, \varphi_2 \}= \frac{{2\sqrt \pi}}{{\Gamma(- \nu)}}$ \cite{susy-article-ref:26}, where $\Gamma(-\nu)$ -- gamma-function. Following the procedure from previous chapter, the general solution of auxiliary equation is chosen in the form: 
	\begin{equation}
		\varphi(\xi, \bar \varepsilon, 1)= D_\nu(\sqrt 2 \xi)+D_\nu(- \sqrt 2 \xi)
	\label{susy-article-eqs:33}
	\end{equation}
As one can see from (\ref{susy-article-eqs:33}) $\varphi(x, \bar \varepsilon, 1)$ is an even function of $\xi$. To obtain the exact form of superpotential, the integral, that enter the definition of the $W(x, \varepsilon, \lambda)$ and normalization constant $N^{-2}$ need to be calculated from (\ref{susy-article-eqs:22}), (\ref{susy-article-eqs:23}). Due to the symmetry of $\varphi(\xi, \bar\varepsilon,1)$, expression of the integral simplifies and obtain the form:
	\begin{equation}
		\displaystyle 1+\lambda N^{- 2}\int\limits_{- \infty}^\xi{\frac{{dt}}{{\left({\varphi_1+\varphi_2}\right)^2}}}= 1 - \frac{{\lambda N^{- 2}}}{{2W\{\varphi_1, \varphi_2 \}}}\left[{\left({\frac{{\varphi_1(\xi, \bar \varepsilon)- \varphi_2(\xi, \bar \varepsilon)}}{{\varphi_1(\xi, \bar \varepsilon)+\varphi_2(\xi, \bar \varepsilon)}}}\right)+\Delta(+\infty, \bar \varepsilon, 1)}\right]
	\label{susy-article-eqs:34}
	\end{equation}
Hence, superpotential could be expressed in the form:
	\begin{equation}
		\displaystyle W\{\xi, \bar \varepsilon, \lambda \}= \ln \left({\frac{{(1+\frac{\lambda}{2}+\frac{\lambda}{{2\Delta(+\infty, \bar \varepsilon, 1)}})\varphi_1+(1+\frac{\lambda}{2}- \frac{\lambda}{{2\Delta(+\infty, \bar \varepsilon, 1)}})\varphi_2}}{{\varphi_1+\varphi_2}}}\right)
	\label{susy-article-eqs:35}
	\end{equation}
	$$
	\displaystyle N^{- 2}= - \frac{{W\{\varphi_1, \varphi_2 \}}}{{\Delta(+\infty, \bar \varepsilon, 1)}}= \frac{{2\sqrt \pi}}{{\Gamma(- \nu)}}
	$$
Using the asymptotic value of the parabolic cylinder function, we have $\Delta(+\infty, \bar\varepsilon, 1)=-1$ and expression for superpotential become quite compact:
	\begin{equation}
		\displaystyle W(\xi, \bar \varepsilon, \lambda)= - \frac{1}{2}\ln \left({\frac{{\varphi_1(\xi, \bar \varepsilon)+(1+\lambda)\varphi_2(\xi, \bar \varepsilon)}}{{\varphi_1(\xi, \bar \varepsilon)+\varphi_2(\xi, \bar \varepsilon)}}}\right)= - \frac{1}{2}\ln \left({\frac{{\varphi(\xi, \bar \varepsilon, 1+\lambda)}}{{\varphi(\xi, \bar \varepsilon, 1)}}}\right)
	\label{susy-article-eqs:36}
	\end{equation}
From (\ref{susy-article-eqs:36}) follows, that value of parameter $\lambda$ is limited by the condition $\lambda>-1$. Using the expressions from second chapter and superpotential (\ref{susy-article-eqs:36}) for $H_-^-$ and $\psi_-^-(x, E)$, we could derive the exact form of Hamiltonians, which form the super Hamiltonian, and corresponding wave functions ($\psi_+^-(\xi, E_i)$ -- wave functions of $HO$):
	\begin{equation}
		\begin{array}{c}
		\displaystyle H_-^- = H_+^- - \frac{{d^2}}{{dx^2}}\ln \left({D_\nu(\sqrt 2 \xi)+D_\nu(- \sqrt 2 \xi)}\right),\\
		\displaystyle \psi_-^-(\xi, E_i)= \frac{1}{{\sqrt{2(E_i - \bar \varepsilon)}}}\frac{{W\left\{{\psi_+^-(\xi, E_i), \varphi(\xi, \bar \varepsilon, 1)}\right\}}}{{\varphi(\xi, \bar \varepsilon, 1)}}\\
		\displaystyle \psi_-^-(x, E = 0)= \frac{{N^{- 1}}}{{\left({D_\nu(\sqrt 2 \xi)+D_\nu(- \sqrt 2 \xi)}\right)}}= \frac{{N^{- 1}}}{{\varphi(x, \bar \varepsilon, 1)}}, \quad \;N^{- 2}= \frac{{2\sqrt \pi}}{{\Gamma(- \nu)}}
		\end{array}
	\label{susy-article-eqs:37}
	\end{equation}
Analysis of the potential (\ref{susy-article-eqs:37}) reveals that several local minima exist only when $0<\bar\varepsilon<\frac{1}{2}$. With values of $\bar\varepsilon$ close to right boundary the third local minimum appears (Fig.\ref{susy-article-fig:1}) and depth of outside minima increases. It should be noted that in the terms of dimensionless variable $\xi$ the only way to vary the form of potential is through varying $\bar\varepsilon$ and $\lambda$. In the case of natural units, additionally, the form of potential(in particular, positions of local minima)could be changed through variation of $\omega$.
Connection between $H_+^+$, $\psi_+^+(\xi, E_n)$ and $H_+^-$, $\psi_+^-(\xi, E_n)$ correspondingly could be treated in the same manner. Using (\ref{susy-article-eqs:36}), the following expressions for $H_+^+$ and $\psi_+^+(\xi, E_n)$ could be obtained from (\ref{susy-article-eqs:37}) by substitution $\varphi(\xi,\bar \varepsilon, 1)\rightarrow\varphi(\xi,\bar \varepsilon, \lambda+1)$:
	\begin{equation}
		\begin{array}{c}
		\displaystyle H_+^+= H_+^- - \frac{{d^2}}{{dx^2}}\ln \left({D_\nu(\sqrt 2 \xi)+(1+\lambda)D_\nu(- \sqrt 2 \xi)}\right),\\
		\\
		\displaystyle \psi_+^+(\xi, E_i)= \frac{1}{{\sqrt{2(E_i - \bar \varepsilon)}}}\frac{{W\left\{{\psi_+^-(\xi, E_i), \varphi(\xi, \bar \varepsilon, \lambda+1)}\right\}}}{{\varphi(\xi, \bar \varepsilon, \lambda+1)}}\\
		\\
		\displaystyle \psi_+^+(\xi, E = 0)= \frac{{N_{\lambda+1}^{- 1}}}{{\left({D_\nu(\sqrt 2 \xi)+(\lambda+1)D_\nu(- \sqrt 2 \xi)}\right)}}= \frac{{N_{\lambda+1}^{- 1}}}{{\varphi(x, \bar \varepsilon, \lambda+1)}},\\
		\\
		\displaystyle N_{\lambda+1}^{- 2}= \frac{{2(\lambda+1)\sqrt \pi}}{{\Gamma(- \nu)}}
		\end{array}
	\label{susy-article-eqs:38}
	\end{equation}
From this relations the restriction $-1< \lambda$ follows. As could be seen from Fig.\ref{susy-article-fig:2}, potential $U_+^+(\xi, \bar\varepsilon, \lambda)$ has well pronounced asymmetry, which increases when $\lambda \to -1$. The value $\bar\varepsilon = 0.47$ corresponds the region, where $U_+^+(\xi, \bar\varepsilon, \lambda)$ has three local minima. It shloud be noted, that depth of central minima also increases with $\lambda\to -1$. Given potentials and corresponding wave functions could useful for research of resonant tunneling phenomenon \cite{susy-article-ref:4}.

\begin{figure}
\begin{center}
 \includegraphics[width=0.5\linewidth]{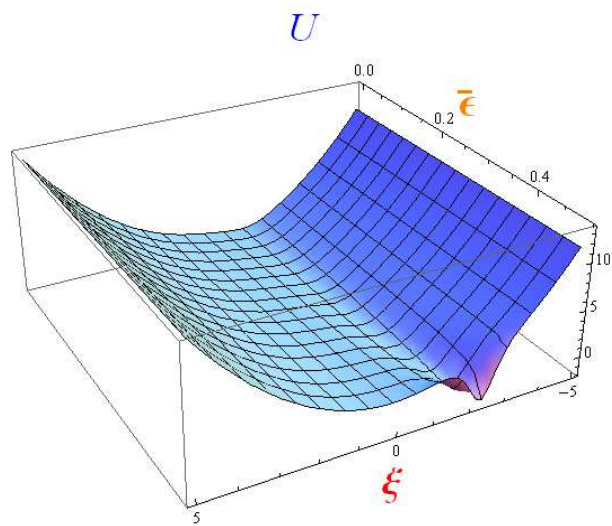}
  \caption{Potential $U_-^-(\xi, \bar \varepsilon)$($0 < \bar \varepsilon < 0.5$).\label{susy-article-fig:3}}
\end{center}
\end{figure}
\textbf{b) Partial supersymmetry breaking}

The utilization of the particular solution of auxiliary equation for derivation of superpotential (\ref{susy-article-eqs:15}) realizes the situation of partial supersymmetry breaking. As it was mentioned above, the spectrum of $H_-^-$ does not contain states with $E = 0$, because their wave function is nonnormalizable. Potential $U_-^-(\xi, \bar\varepsilon)$ is presented on Fig.\ref{susy-article-fig:3}. At the same time, zero state appears in the spectrum of $H_+^+$ with wave function:
	\begin{equation}
		\psi_+^+(\xi, E = 0)= \frac{{N_\lambda^{- 1}}}{{\varphi(\xi, \bar \varepsilon, \lambda)}}= \frac{{N_\lambda^{- 1}}}{{\left({\varphi_1(\xi, \bar \varepsilon)+\lambda \varphi_2(\xi, \bar \varepsilon)}\right)}}
	\label{susy-article-eqs:39}
	\end{equation}

Normalization constant $N_{\lambda}^{-2}$ could be calculated using (\ref{susy-article-eqs:35}) and has the form $\displaystyle N_\lambda^{-2}=\lambda W\{\varphi_1, \varphi_2\}= \frac{{2\lambda \sqrt \pi}}{{\Gamma(- \nu)}}$. Thus, the limitation is $\lambda > 0$. Hamiltonian has the form $\displaystyle H_+^+=H_+^- - \frac{{d^2}}{{d\xi^2}}\ln\left({\varphi(\xi, \bar\varepsilon, \lambda}\right)$. Potential ($\bar\varepsilon = 0.47$) is presented in Fig.\ref{susy-article-fig:2} at $\lambda > 0$.

When $\displaystyle \bar \varepsilon = - \frac{1}{2}$ ($D_{- 1}(\sqrt 2 \xi)= e^{\frac{{\xi^2}}{2}}\sqrt{\frac{\pi}{2}}\left({1 - \Phi(\xi)}\right)$, where $\Phi(\xi)$ -- error function), potentials $U_-^-(x, \bar\varepsilon)$ corresponds the potential, given in \cite{susy-article-ref:27} (the only difference is additional constant term).
	\begin{equation}
		U_-^-\left(\xi, \bar \varepsilon = - \frac{1}{2}\right)= \left({\frac{{\xi^2 - 1}}{2}}\right)+\frac{4}{{\sqrt \pi}}\, \frac{{e^{- \xi^2}}}{{\left({1 - \Phi(\xi)}\right)}}\left[{\frac{{\;e^{- \xi^2}}}{{\sqrt \pi \left({1 - \Phi(\xi)}\right)}}- \xi}\right]
	\label{susy-article-eqs:310}
	\end{equation}
This potential is single-well, as well as corresponding $\displaystyle U_+^+(\xi, \bar\varepsilon = -1/2, \lambda)$, because existence of several local minima is possible only when $\displaystyle 0 < \bar \varepsilon < \frac{1}{2}$. Using the form-invariance property for $U_+^+$:
	\begin{equation}
		U_{+}^+(\xi, \bar \varepsilon = -{\raise0.7ex\hbox{$1$}\!\mathord{\left/ 
		{\vphantom{1 2}}\right.\kern-\nulldelimiterspace}
		\!\lower0.7ex\hbox{$2$}}, \lambda)= \frac{{\xi^2 - 1}}{2}- \frac{{d^2}}{{d\xi^2}}\ln \left({(1+\lambda)-(1 - \lambda)\Phi(\xi)}\right)
	\label{susy-article-eqs:311}
	\end{equation}
Spectrum of Hamiltonian with potential (\ref{susy-article-eqs:311}) contains, unlike (\ref{susy-article-eqs:310}), the state with $E = 0$ and wave function of the form:
	\begin{equation}
		\displaystyle \psi_+^+(\xi, E = 0)= \frac{{\left({2\lambda \sqrt \pi}\right)^{1/2}e^{- \frac{{\xi^2}}{2}}}}{{\left({(1+\lambda)-(1 - \lambda)\Phi(\xi)}\right)}}
	\end{equation}
 
\section{Conclusions}

The presented paper is devoted to construction and study of general properties of Hamiltonians with multi-well potentials in the framework of $N = 4~SUSY~QM$. The emphasis is made on the research of the distinctive features of these Hamiltonians without concretization of the form of initial Hamiltonian. The given relation for certain type of integrals, containing the fundamental solutions of the Shr\"odinger type equations, allows to show the form-invariance of the isospectral Hamiltonians with multi-well potentials, obtained in the framework of $N = 4~SUSY~QM$. I.e, with the identical spectra of $H_-^-$ and $H_+^+$, corresponding potentials could be obtained from each other through the substitution of parameters. Moreover, this relation allows to analytically calculate normalization constants of zero modes, using only asymptotic values of fundamental solutions. Procedure of construction of isospectral Hamiltonians in the framework of $N = 4~SUSY~QM$ also allows the partial symmetry breaking.
On the model of harmonic oscillator we derive the exact form of isospectral Hamiltonians with multi-well potentials and corresponding wave functions. The existence of the parameters allows to vary the form of potential in the wide range. This is especially important in research of the phenomena such as tunneling, which are sensitive to the structure of the multi-well potentials. Moreover, the derived expressions could serve as a basis for construction of interesting exactly solvable models of stochastic processes, Kramers problem, stochastic resonance, to mention only few.

Author is grateful to M.Plyushchay for valuable advices and appreciate M.Konchatnij and G.Ivashkevych for help with paper preparation.

\end{document}